\documentclass[twocolumn,showpacs,preprintnumbers,amsmath,amssymb]{revtex4}
\input epsf.sty

\usepackage{graphicx}
\usepackage{dcolumn}
\usepackage{bm}

\begin{document}


\title{Extrapolation method in shell model calculations with deformed basis}

\author{Takahiro Mizusaki}

\affiliation{
Institute of Natural Sciences, Senshu University, Kawasaki, Kanagawa, 214-8580, Japan 
}

\date{\today}

\begin{abstract}
An extrapolation method in shell model calculations with deformed basis is presented,
which uses a scaling property of energy and energy variance for a series of
systematically  approximated wave functions to the true one.
Such approximated wave functions are given by variation-after-projection method
concerning the full angular momentum projection.  
This extrapolation needs energy variance, which amounts to the calculation of expectation value 
of square of Hamiltonian $\hat{H}^2$. 
We present the method to evaluate this matrix element and show that large reduction 
of its numerical computation can be done by taking an advantage of time-reversal symmetry. 
The numerical tests are presented for $fp$ shell calculations with a realistic residual
interaction.
\end{abstract}

\maketitle

Extrapolation is an intriguing and useful concept in computational physics in
quantum many-body problems in condensed matter physics, nuclear structure physics
and other fields.
The Hilbert spaces of quantum many-body systems can be finite but essentially large 
or infinite.
To handle such Hilbert spaces completely by numerical methods,
there exist various difficulties.
However if we can handle its dominant but small subspace,
the contribution of the rest huge subspace can be taken into account by extrapolation.
More concretely we can estimate the exact energy by extrapolation if we know 
a scaling property 
how to change energy as a function of a certain physical quantity.

In the nuclear structure physics, shell model is one of fundamental frameworks.
For shell model calculations, one needs handle very huge Hilbert 
space. One of conventional approaches is diagonalization with spherical single particle basis states, 
which recently have been applicable to quite large-scale problem with dimension up to $10^9$\cite{caurier}.
However, for further larger spaces, diagonalization is impossible. 
To overcome such a limitation of diagonalization, various methods \cite{smmc,dmrg,vampir,qmcd1,qmcd-rev,
txp,ecm,papenbrock,andreozzi,ncsm,xp,xp1} have been proposed and have been developed.
Among them, extrapolation methods \cite{txp,ecm,papenbrock,andreozzi,ncsm,xp} have been developed, combined with
diagonalization of the Hamiltonian matrix evaluated with spherical basis states. 
For instance, exponential convergence method (ECM) 
\cite{ecm}
can predict the exact energy by assuming exponential behavior of energy as a 
function of the number of basis states and empirical successes of its assumption
have been reported with various truncation schemes \cite{ecm,papenbrock,andreozzi}.

We have also proposed an extrapolation method in shell model calculations,
combined with Lanczos diagonalization for a series of truncated spaces \cite{xp,xp1},
where we take an advantage of a scaling property between 
the energy difference $\delta E$ and energy variance $\Delta E$ for a series 
of systematically approximated wave functions. 
Such a scaling property 
has been originally introduced \cite{imada1,sorella} in the study of Hubbard models.
We have introduced this scaling property into shell model calculations \cite{xp}.
Moreover, we have shown that, for precise estimation, higher order term of expansion
can be taken into consideration \cite{xp1}.
Very recently this scaling property has been successfully applied to the no core shell model \cite{ncsm}.

Thus two kinds of extrapolation methods \cite{ecm,papenbrock,andreozzi,ncsm,xp,xp1} 
have been proposed and are quite promising.
However they have a common drawback that they strongly depend on 
the diagonalization with spherical basis states.
The dimension of necessary truncated subspace heavily 
depends on each nucleus as the dimension of whole shell model space
does. 

In the present paper, we propose an extrapolation method in the shell model calculations
with deformed bases, which are expected to be suited for deformation of nuclei. 
As our extrapolation scheme depends on only energy and energy variance, 
we can apply it to deformed basis. 
We will show that the exact energy can be extrapolated 
by the behavior of energy evaluated by single Slater determinant with angular momentum projection. 
 
First we briefly summarize our extrapolation scheme.
We consider a series of systematically approximated wave functions as
$\left| {\psi _1 } \right\rangle $, $\left| {\psi _2 } \right\rangle $,$\cdots$.
Here we assume that they are systematically generated and their energies becomes 
lower as $E_1 \ge E_2 \ge \cdots $ where $E_i$ is an energy of $\left| {\psi _i } \right\rangle $.
Later we will discuss how to prepare them in deformed basis in detail.
The approximated wave function contains large fraction of true wave 
function and contamination of rest wave function is small. 
By expanding this small value, a following relation is obtained.
We define the difference $\delta E$ between the lowest energy eigenvalue 
$\langle  {\hat H}  \rangle$ of approximated wave functions 
and true energy eigenvalue $\langle {\hat H} \rangle _0$ as
$\delta E = \langle {\hat H} \rangle -\langle {\hat H} \rangle_0$.
The energy variance  $\Delta E$ of approximated wave function is also defined as
$
\Delta E = \frac{{\left\langle {\hat H^2 } \right\rangle  - \left\langle {\hat H} \right\rangle ^2 }}{{\left\langle {\hat H} \right\rangle ^2 }}
$.
As a first order approximation, we can show a proportionality as $\delta E \propto \Delta E$ \cite{xp}
by expanding $\delta E$ as a function of $\Delta E$.
We can take into account second order effects for precise estimation \cite{xp1}.
By these scaling properties, we can estimate the exact energy by taking 
$\Delta E \to 0$.  

Next we will discuss how to prepare such systematically approximated 
wave functions in deformed basis.
We consider an angular momentum projected deformed wave function
such as 
\begin{equation}
\left| \psi_{J,M} \right \rangle =  \sum_K g_K P_{MK}^J \left| \psi  \right\rangle 
\end{equation}
where $ P_{MK}^J $ is angular momentum projection operator and $g$'s are coefficients.
The wave function $| \psi \rangle$ is a Slater determinant as
$ \left| \psi  \right\rangle  = \prod {a_\alpha ^\dag  } \left| 0 \right\rangle $
where $\left| 0 \right\rangle $ is a vacuum state and
$a_\alpha ^\dag$'s are creation operators on deformed orbit $\alpha$ and are
defined as 
$a_\alpha ^\dag   = \sum\limits_i {D_{\alpha ,i} c_i^\dag  } $.
The $ c_i^\dag$ denotes creation operator of spherical orbit $i$ and $D$'s are coefficients
and are determined by the following variational equation:
\begin{equation}
\delta \left\{ {\frac{{\left\langle \psi_{J,M}  \right|\hat{H} \left| \psi_{J,M}  \right\rangle }}
{{\langle \psi_{J,M}  \left| \psi_{J,M}  \right\rangle }}} \right\} = 0.
\label{phf}
\end{equation}
In order to solve above equation, we minimize the energy expectation value
with angular momentum projected wave function
concerning $D_{\alpha ,i} $ \cite{ring}.
This can  be achieved by gradient method where we evaluate the gradient vector 
in the projected energy surface and then we change the wave function along the steepest 
descent line. By iterating this procedure, the variation of the angular momentum projected energy is achieved.
In general, obtained wave function in this procedure corresponds to local minimum in angular momentum projected energy surface.
By examining several local minima, we can obtain lowest minimum in the practical calculations.
Here we assume that wave function obtained in this procedure is a good approximation
for the true wave function, that is, it has a large overlap to true one.
In practical shell model calculations, this assumption seems to be good \cite{fda}.

Under this assumption, we consider how to construct a series of systematically
approximated wave functions. The wave function of lowest minimum
is written by $\left| {\varphi _{\min } } \right\rangle $.
The shell model orbits are grouped into upper and lower orbits, 
for instance, in the $fp$-shell, 
four orbits are grouped by \{$f_{7/2}$\} and \{$p_{3/2},f_{5/2},p_{1/2}$\}.
By this grouping, we consider a following parameterized projected wave functions:
\begin{equation}
\left| \varphi_{J,M} \left( x \right) \right \rangle =  
\sum_K g_K P_{MK}^J \left| \varphi \left( x \right) \right\rangle 
\end{equation}
and
\begin{equation}
\left| {\varphi \left( x \right)} \right\rangle  = \prod {a_\alpha ^\dag  \left( x \right)} \left| 0 \right\rangle 
\end{equation}
where the creation operator $a_\alpha ^\dag  (x)$'s are defined as
$a_\alpha ^\dag  (x) = \sum\limits_i {D_{\alpha ,i}^{(\min )} x_i c_i^\dag  } $.
The  $D_{\alpha ,i}^{(\min )}$ corresponds to the one of  $\left| {\varphi _{\min } } \right\rangle $ 
and $x_i$ takes $1$ or $x$, depending on the group of $i$. 
The $g$'s are determined at each $x$.
When the $x$ increases or decreases from unity, the angular momentum projected energy of
$\left| {\varphi \left( x \right)} \right\rangle$ increases
because of 
$
\left| {\varphi \left( {x = 1} \right)} \right\rangle  = \left| {\varphi _{\min } } \right\rangle 
$.
In this way, we can systematically generate a series of well-approximated 
wave functions near the lowest minimum continuously as a function of energy.
This construction is not unique and we may find other methods.
However this way to construct a series of wave functions works well 
for $fp$-shell calculations as we will show later. 

Note that, in general, we have no method to know whether or not a 
considered wave function is well-approximated to the true wave function
unless information of the true state is available.
Empirically  an optimized projected wave function by variation after angular momentum projection is known to describe deformed state well while we can not, strictly, 
know the precision of its approximation.
In the present method, the energy variance can give an index of precision of approximation.

For the extrapolation by these deformed bases,
expectation value of $\langle \hat{H}^2 \rangle$ is necessary. As we consider a realistic description,
we use a general two-body shell model effective interaction as,
\begin{equation}
\hat{H} = \sum\limits_\alpha  {\varepsilon _\alpha  } c_\alpha ^\dag  c_\alpha   + \sum\limits_{\alpha  < \beta ,\gamma  < \delta } {v_{\alpha \beta \gamma \delta } } c_\alpha ^\dag  c_\beta ^\dag  c_\delta  c_\gamma  
\end{equation}
where $\varepsilon$'s and $v$'s are single particle energies and two-body matrix elements, respectively.
Then $\hat{H}^2$ includes one-body, two-body, three-body and four-body terms.
For instance, normal ordered form of four body term is shown by
\begin{equation}
\hat{H}_4  = \sum\limits_{i < j < k < l,\alpha  < \beta  < \gamma  < \delta } {\tilde v_{ijkl\alpha \beta \gamma \delta } c_i^\dag  c_j^\dag  c_k ^\dag  c_l ^\dag 
 c_\delta c_\gamma c_\beta  c_\alpha  } 
\end{equation}
where $\tilde v$ can be given by summing up relevant $v$'s
and is totally antisymmetric for $i,j,k,l$ and for $\alpha,\beta,\gamma,\delta$.
Expectation value of this operator by angular momentum projected state can be
carried out by Wick's theorem. 
We define generalized one-body density matrix as   
\begin{equation}
\rho _{\beta \alpha }  = {{\left\langle \varphi  \right|c_\alpha ^\dag  c_\beta  \left| \psi  \right\rangle } \mathord{\left/
 {\vphantom {{\left\langle \varphi  \right|c_\alpha ^\dag  c_\beta  \left| \psi  \right\rangle } {\left\langle \varphi  \right|\left. \psi  \right\rangle }}} \right.
 \kern-\nulldelimiterspace} {\left\langle \varphi  \right|\left. \psi  \right\rangle }}
\end{equation}
where $\left| \varphi  \right\rangle $ and $\left| \psi  \right\rangle $
are deformed Slater determinants.
For angular momentum projection, density matrices between different wave functions are
necessary.
By this one-body density matrix, we can define two-body density matrix as 
\begin{equation}
\rho _{\alpha \beta \gamma \delta }  = \frac{{\left\langle \varphi  \right|c_\alpha ^\dag  c_\beta ^\dag  c_\delta  c_\gamma  \left| \psi  \right\rangle }}{{\left\langle \varphi  \right|\left. \psi  \right\rangle }} =
 \rho _{\gamma \alpha } \rho _{\delta \beta } -
 \rho _{\delta \alpha } \rho _{\gamma \beta }  
 .
\end{equation}

By the two-body density matrix, we can show the expectation value of the
four-body term $\hat{H}_4$ which can be given by six terms of
product of two two-body density matrices.
As we consider proton-neutron system and isospin projection is not introduced,
two-protons and two-neutrons creation and annihilation part
$c_{\pi ,i}^\dag  c_{\pi ,j}^\dag  c_{\nu ,k}^\dag  c_{\nu ,l}^\dag  c_{\nu ,\delta } c_{\nu ,\gamma } c_{\pi ,\beta } c_{\pi ,\alpha } 
$, which is  
a dominant part of the $\langle \hat{ H}^2 \rangle $ calculation, 
needs only one product of two two-body density matrices.   
Then such a normal ordered form of the $\hat{H}^2$ operator is most efficient for 
numerical calculation.

Next we consider how to reduce computational time by taking an advantage of time-reversal symmetry. By solving eq. (\ref{phf}) for $J=0$ ground state, we can obtain
the wave function with time-reversal symmetry.
For such a case, it is known that an integral domain for angular momentum projection 
can be reduced. 
The projected matrix element of the Hamiltonian $H_{KK'}^J$ is defined by
$
H_{KK'}^J  = \left\langle \psi  \right|\hat{H}\hat{P}_{KK'}^J \left| \psi  \right\rangle 
$,
which is evaluated by a following integral:
\begin{equation}
H_{KK'}^J  = {\textstyle{{2J + 1} \over {8\pi ^2 }}}\int_0^\pi  {d\alpha } \int_0^{{\raise0.5ex\hbox{$\scriptstyle \pi $}
\kern-0.1em/\kern-0.15em
\lower0.25ex\hbox{$\scriptstyle 2$}}} {d\beta } \int_0^{2\pi } {d\gamma } W_{K,K'}^J \left( \Omega  \right)
.
\end{equation}
The $\Omega$ stands for the Euler's angles. The $W_{K,K'}^J \left( \Omega  \right)$
is a sum of relevant product of the Wigner's D-function and rotated  matrix element
$
\left\langle \psi  \right|\hat{H}\hat{R}\left( \Omega  \right)\left| \psi  \right\rangle 
$ where $\hat{R}\left( \Omega  \right)$ is a rotating operator.
Detailed form is written in Ref. \cite{enami}, for instance. 
Due to the reduced integral domain and relations of rotated matrix element concerning 
Euler's angles, computational time reduces $1/8$ shorter.

Furthermore, to reduce amount of computation, as a series of 
systematically approximated 
wave functions for the extrapolation of excited states, we use the same series of ground state. In this case, we can use the same rotated matrix elements for 
$\langle \hat{H}^2 \rangle$ calculation. Therefore we can simultaneously obtain 
extrapolations of several low-lying states in addition to the one of
the ground state. 

\begin{figure}[h]
\begin{picture}(200,200)
    \put(0,0){\epsfxsize 190pt \epsfbox{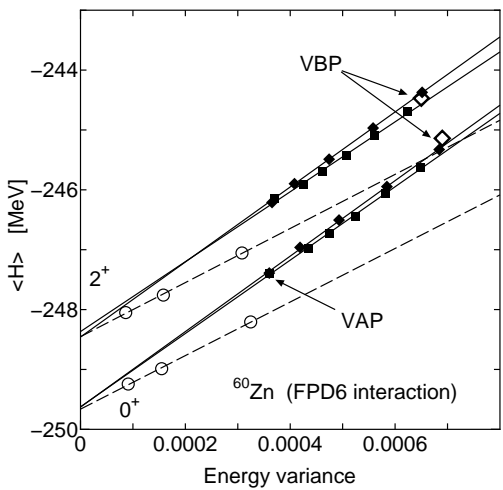}}
\end{picture}
\caption{Extrapolations of two kinds of extrapolation method are shown
for the ground and first excited states of $^{60}$Zn with the FPD6 interaction
\protect\cite{fpd6}.
The open circles are shown for energies and energy variances of 
$t=3 \sim 5$ truncated spaces. Their extrapolations are shown by dashed lines.
The filled squares and filled diamonds are  energies and energy variances of deformed states with 
angular momentum projection for groupings (i) and (ii), respectively. 
The angular momentum projected energies and energy variances for $J=0$ and 2
from the Hartree Fock wave function are shown by open diamonds
and are indicated by the word VBP.
The energy of variation after angular momentum projection for $J=0$ state 
is labeled by the word  VAP. 
}
\end{figure}

In order to test the present extrapolation method, we take an example of 
$^{60}$Zn in the $fp$ shell model space, the dimension of which Hilbert space
is largest in the $fp$ shell model space and is 2,292,604,744 in $M=0$ space. 
We use the FPD6 effective interaction \cite{fpd6}.
To test the present extrapolation, the exact ground state energy is needed. 
The exact ground state energy of this nucleus may be obtained by the state-of-the-art 
Lanczos shell model diagonalization if we take quite long computational time.
However here we use extrapolation method which we have proposed in Refs. \cite{xp,xp1}.
The extrapolation method also use the scaling property of energy and 
energy variance while we combined it to Lanczos shell model diagonalization.
By the minimum number $t$ of the holes in the $f_{7/2}$ orbit, we can systematically
define the truncation spaces as $\oplus_t f_{7/2}^{(16-t)} r^{4+t} $ 
($r=p_{3/2}, f_{5/2}, p_{1/2}$). Truncation spaces are labeled by $t$.
As the $t$ increases, its truncation space becomes larger and the lowest energy state
$\left| {\varphi _t } \right\rangle $
of such a truncated space becomes a better approximation to the true state.

In Fig. 1, we use the truncation spaces with $t=$3, 4 and 5.
The dimensions of truncated shell model spaces are 4794006, 31077402 and 133533398 respectively. 
For each space, we diagonalize the shell model Hamiltonian
matrix and obtain the approximated wave functions of the true ground state.
Then we evaluate the matrix element of $\hat{H}^2$.
The extrapolation is shown in Fig. 1.  
For the diagonalization in truncated spaces, we
use the shell model code MSHELL\cite{mshell}. 

As deformed basis, Hartree-Fock energy is primarily important and is 
-243.03 MeV for $^{60}$Zn. 
First step of improvement of wave function is the angular momentum projection.
The angular momentum projected energies for $J=0$ and 2 states from the Hartree Fock 
wave function are shown in Fig. 1. 
Angular momentum projection improves energy and we can explicitly handle
angular momentum quantum numbers. This approach is known to be the VBP
(variation before projection). 
Next step is a variation after  angular momentum projection.
We can determine the minimum energy in the angular momentum projected 
energy surface for $J=0$. This energy is also shown in Fig. 1.
By these energies and energy variances, we may extrapolate the true energy.
In this case, this extrapolation is not so bad. However, relation between
these two wave functions is not clear and extrapolation from only two data 
is not reliable.

For extrapolation, a series of systematically approximated wave functions is important.
To construct such a series, four $fp$ shell orbits are
grouped in following two ways: (i) \{$f_{7/2},p_{3/2}$\} and \{$f_{5/2},p_{1/2}$\}
and (ii) \{$f_{7/2}$\} and \{$p_{3/2}, f_{5/2},p_{1/2}$\}.
For these groupings, we change $x$ parameter in $\left| {\varphi \left( x \right)} \right\rangle $. The dependence of $x$ in  $a_\alpha ^\dag  (x)$ is simple, but
for the angular momentum projected energy, the dependence of $x$ is no more simple
and is non-linear. However by rescaling the energy as a function of energy variance,
we can obtain a simple scaling again. 
In Fig. 1, we take $x$=1.0, 0.8, 0.75, 0.7, 0.65 and 0.6 for grouping (i)
and $x$=1.0, 0.8, 0.7, 0.6, and 0.5 for grouping (ii).
These data of energy to the energy variance for ground and excited states
are well-aligned and we can find fine  proportionalities, by which  
we can make sure extrapolation.
The energies of the present extrapolation method agree with
those of the previous method within 0.1 MeV. 
In the respect of the computational amount, previous extrapolation 
with spherical basis \cite{xp,xp1} needs four days while this new 
approach needs less than half day in this calculation.  
Therefore the present extrapolation method is superior to the previous one.

In the present method, we solve eq. (\ref{phf}) for $J=0$ ground state while 
extrapolations are carried out for several low-lying states from the 
series of wave functions for $J=0$ state. 
This method has an advantage for drastically reducing computational time.
However for higher spin states, the approximation may become worse.
Therefore, we test what extent the extrapolation
works well for several low-lying yrast states by this approach. 
Here we take  $^{48}$Cr as an example.  
For comparison, we can give exact energies for yrast states by Lanczos 
diagonalization.  
In the left panel of Fig. 2, we show the exact level scheme.
In the right panel, the energies as a function of energy variances
for $x$=1.0, 1.1, 1.2 and 1.3 are plotted.
Four $fp$ shell orbits are
grouped in a following way: \{$f_{7/2}$\} and \{$p_{3/2},f_{5/2},p_{1/2}$\}.
Because $x=1.0$ corresponds to the minimum energy, energies with $x > 1$
are higher than minimum energy.
By this construction, for a series of wave functions  $x \ge 1$ ($x \le 1$),
excitation from the $f_{7/2}$ orbit to other upper three orbits
increases (decreases).  As more excitation is needed, in general, 
to generate high spin states, we consider a series for $x \ge 1$.
In Fig. 2, 
the extrapolated energies agree with the exact ones within 0.1 MeV,
which means that a series of wave functions for $J=0$ state works 
for other low-lying states.  
However, minimum of variance for higher spin state increases.
This means deterioration of approximation, while such deterioration  is somewhat 
compensated by the extrapolation procedure.
Thus the present approach works well for low spin yrast states.

For higher spin, deterioration of approximation is evident.
This point is expected to be improved by optimizing a deformed wave 
function for each spin, which needs longer computational time.
Moreover, here we show energies for $J\le8$ states while this nucleus
is known to have a backbending \cite{cr48}. 
For such a large change of structure in wave functions as a function of spin,
such an optimization becomes necessary.

\begin{figure}[h]
\begin{picture}(200,200)
    \put(0,0){\epsfxsize 190pt \epsfbox{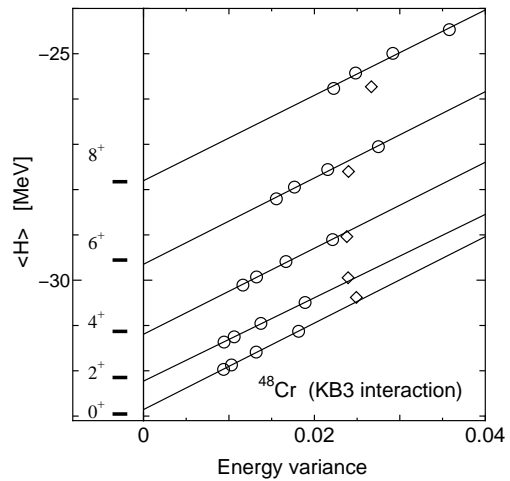}}
\end{picture}
\caption{ Left panel: The exact level scheme is shown.
Right panel: Extrapolations of the present method are shown
for several low-lying excited states ($0^+$, $2^+$, $4^+$, $6^+$ and $8^+$  
) of $^{48}$Cr with the KB3 interaction
\protect\cite{kb3}.
The angular momentum projected energies and energy variances  
from the Hartree Fock wave function are shown by open diamonds. }
\end{figure}

Finally we discuss an advantage of the present extrapolation method to other extrapolation methods.
In an open-shell nuclei, deformed states often appear due to residual interaction.
To describe deformed states by spherical bases, we need almost entire 
spherical bases. Consequently truncation is not so effective and
the convergence of energy as a function of the number of spherical 
basis state becomes very slow.  According to the ECM\cite{ecm}, necessary number of spherical 
basis state increases exponentially as we need more precise energy.
On the other hand, it is evident that deformed basis can handle deformed solution more naturally.
In the present paper, we use always one Slater determinant.
This fact is in sharp contrast with other extrapolation methods
and becomes a distinct advantage for larger shell model problems, 
where a physically relevant truncation by spherical basis 
becomes severely difficult.
Another merit of the present method is to handle all  nucleus  in the 
considered model space within almost the same amount of computation. 
In extrapolation methods based on the spherical shell model,
as the number of valence nucleons reaches half of degeneracy of orbits,
amount of computation rapidly increases. Therefore feasibility of
the present method is wider than other extrapolation methods.  
Moreover the present method has a close link to  mean-field or projected 
mean-field theory. It is also an advantage that physical interpretations
of the obtained results based on mean-field is easy.

In summary, we have proposed an extrapolation method in shell model calculations
by deformed wave functions with angular momentum projection.
Our extrapolation technique relies on a scaling property between energy and 
energy variance of a series of systematically approximated wave functions. 
To apply it to shell model calculations, a key issue is how to construct such a 
series.
In our previous papers \cite{xp,xp1}, we consider physically relevant truncation subspaces
and we solve shell model problem within these subspaces by Lanczos method.
Then we obtain such a series and apply the extrapolation technique.
On the other hand, in the present paper, we consider a deformed wave functions
with angular momentum projection.
By variation-after-projection, we can solve shell model problem under a
restriction of single Slater determinant with angular momentum projection. 
This wave function often becomes a relevant approximation to the true state,
especially when nucleus is deformed.
We have shown that  a series of approximated wave functions
can be generated  by changing their structure  systematically.
To evaluate the energy variances by such an angular momentum projected Slater determinant, the Wick's theorem is used.  
As we do not use isospin projection for proton-neutron systems, 
most time consuming part of  $\langle \hat{H}^2 \rangle$ calculations is rather
simplified.
Time reversal symmetry is also useful to reduce the numerical calculations
concerning angular momentum projection.  
By these developments, we have shown that the extrapolation method 
with deformed basis works well.

In the present paper, for minimizing necessary amount of computation,
we introduced several restrictions. However, 
for more precise extrapolation and description of
excited states with the same quantum numbers, we can introduce a 
diagonalizaiton with several optimized Slater determinants
and optimization of deformed bases for each spin.
Such important extensions will be able to be realized with a help of parallel 
computation.   
 
As applications of the present method, $fpg$ shell is important,
whose orbits are $p_{3/2}$, $f_{5/2}$, $p_{3/2}$ and $g_{9/2}$.
The ground state of $A \sim 80$ nuclei is determined by a subtle competition 
between oblate and prolate deformed states and occupation of the $g_{9/2}$ orbit is essential,
which means that relevant truncation spaces become quite huge. 
On the other hand, deformed states can be well-approximated by well-optimized Slater 
determinant with angular momentum projection.  The present extrapolation method
can handle such a subtle competition.
Moreover, the present extrapolation may also apply to the study of fractional quantum Hall
effect formulated on the Haldane sphere \cite{fqhe}.  Such studies are being in progress. 

This work was supported in part 
by Grant-in-Aid for Specially 
Promoted Research (13002001) from the Ministry of Education, Science and Culture. 
This work was also supported by the Grant-in-Aid of the Promotion 
and Mutual Aid Corporation for Private Universities of Japan at 2003 and 2004.

\end{document}